\documentclass{aa}
\usepackage{natbib}
\bibpunct{(}{)}{;}{a}{}{,}
\usepackage{graphicx}
\usepackage{times}

\begin{document}

\title{Wide field weak lensing observations of A1689}

\author{D. Clowe \inst{1,2} \and P. Schneider \inst{1,2} }

\offprints{D. Clowe}

\institute{Institut f\"ur Astrophysik und Extraterrestrische Forschung der 
Universit\"at Bonn, Auf dem H\"ugel 71, 53121 Bonn, Germany \and 
Max-Planck-Institut f\"ur Astrophysik, Karl Schwarzschild Str. 1, 85741
Garching, Germany}

\date{Received 11 June 2001 / Accepted 21 September 2001}

\abstract{
We present a mass profile for A1689 from $0.13 h^{-1} \mathrm{Mpc} < r < 2 h^{-1} \mathrm{Mpc}$ from a weak lensing analysis of a 
$34\arcmin \times 34\arcmin$ $R$-band image from the 
ESO/MPG Wide Field Imager.  We detect the gravitational shearing of a 
$23<R<25.5$ background galaxy population even at the edge of the image with a 
$4\sigma $ significance, and find a two-dimensional mass reconstruction has
a $13.4\sigma$ significance mass peak centered on the brightest cluster
galaxy.  This peak is well fit by both a $1028\pm 35$ km/s singular isothermal
sphere and a $r_{200} = 1.28 \:\mathrm{Mpc}, c = 6$ ``universal'' CDM profile, although
the ``universal'' CDM profile provides a better fit with 95.5\%  confidence.
These mass measurements are lower than most of those derived by other means
and we discuss possible reasons for weak lensing providing an underestimate
of the true mass of the cluster.  We find that the correction factors needed
to reconcile the weak lensing mass models with the strong lensing Einstein
radius would result is a much larger fraction of faint stars and foreground
and cluster dwarf galaxies in the $23<R<25.5$ object catalog than is seen in 
other fields.
\keywords{Gravitational lensing --
          Galaxies: clusters: individual: A1689 --
          dark matter}
}

\maketitle

\section{Introduction}
Clusters of galaxies provide several methods of constraining cosmological
and dark matter models.  Two of the most important properties of the clusters
in constraining these models are the total mass and the mass profile.
The three most common techniques in measuring the masses of clusters are
dynamical measurements from the line-of-sight velocity dispersion of cluster
galaxies, the X-ray luminosity and temperature of intra-cluster gas, and
measurement of the distortion of background galaxy images by the cluster potential.
In recent years, the dispersion of the masses measured by the various
techniques has been reduced by improved data and the reduction of systematic
errors, so that in many clusters the mass measurements are in good agreement.
There are a few clusters, however, in which the three mass measurement
techniques are not in agreement, one of the most striking of which is 
\object{Abell 1689}.

\object{Abell 1689} is one of the richest clusters ($R=4$) in the Abell catalog, but
the number counts may be enhanced by a number of lower redshift groups in the
field \citep{TCG90.1}.  A velocity dispersion of 2355 
$^{+238}_{-183}$ km/s has been measured from 66 cluster members 
\citep{TCG90.1}, although a more conservative criteria for cluster membership
reduced the velocity dispersion to 1848$\pm 166$ km/s \citep{GU89.1}.  By
using a technique which identifies substructure from both redshift and
positional information, \citet{GI97.1} divide the cluster galaxies into
three distinct groups each with velocity dispersions of 250-400 km/s, which
would coadd to a mass equivalent to an $\approx 560$ km/s isothermal sphere.

A set of giant arcs $\sim 45\arcsec $ from the brightest cluster galaxy
(hereafter BCG) provide a best-fit model of a 1400 km/s isothermal sphere
centered on the BCG and a 700 km/s isothermal sphere located 1$\arcmin $
northeast of the BCG \citep{MI95.2}.  Analysis of the
depletion of background galaxies around the cluster core, due to the 
deflection and magnification of the galaxies by the gravitational potential, 
results in a best-fit velocity dispersion of 2200$\pm$500 km/s, but is also 
well fit by the double isothermal sphere model from strong gravitational 
lensing \citep{TA98.1}.  An analysis of the change in the background
galaxy luminosity function has measured a mass of $(0.48\pm 0.16) \times
10^{15} h^{-1} \mathrm{M}_{\odot}$ at a $0.25 h^{-1} \mathrm{Mpc}$ radius
from the cluster core, which is consistent with the depletion mass model
\citep{DY01.1}.  Weak lensing shear analysis of the cluster core
gives a mass estimate similar to that of the strong lensing \citep{KA96.1}, 
but suggests that the mass profile is best fit by a power law with index 
$n = 1.4\pm 0.2$ \citep{TY95.1}.  X-ray observations give a
gas temperature of 9-10 keV and best-fit velocity dispersions of 1000-1400
km/s \citep{AL98.1}.

A standard model used to explain these observations involves one or more 
structures (second cluster, filament extending along the line of sight, etc)
located at a redshift only slightly larger than the cluster \citep{TA98.1}.
This would result in line of sight dynamical velocity dispersion
measurement larger than the actual velocity dispersion of the cluster due to
the misinterpretation of cosmological redshift to be velocity.  It would
also result in the lensing mass measurement being higher than the mass of
the cluster core as the lensing measures the total mass surface density
of all the structures along the line of sight.  The X-ray measurements would
thus provide the best estimate of the true mass of the cluster core, as the
small mass structures would provide only a small perturbation to the cluster
X-ray luminosity and gas temperature.  If, however, the additional structure
is undergoing a merger event with the cluster then the gravitational
interaction could create shock heating of the gas and render invalid the
assumptions about isothermality and hydrostatic equilibria on which the
X-ray models are based.
Also, if the \citet{TY95.1} result of the mass profile of the cluster
being steeper than that of an isothermal sphere is correct, then the models 
which convert the dynamical line-of-sight velocity dispersion and X-ray 
measurements to mass estimates are incorrect.   This result, however, is
based on a relatively small field and is therefore uncertain due to the mass
sheet degeneracy and the breakdown of the weak lensing approximation near the
cluster core.  

Due to the new wide-field mosaic CCDs \citep{LU98.1} it is
now possible to obtain a weak lensing signal from low-to-medium redshift
clusters out to large radii ($\sim 2 h^{-1}$ Mpc) in only a few
hours of telescope time.  Here we report on observations of \object{A1689} made with
the Wide Field Imager on the ESO/MPG 2.2m telescope on La Silla.
In Section 2 we discuss the image reduction and object catalog generation.
We analyze the weak lensing signal in Section 3.  Finally, in Section 4 we
compare our mass estimates to others and present our conclusions.  Unless
otherwise stated we assume a $\Omega _\mathrm{m} = 0.3, \Lambda = 0.7$ cosmology
and $H_0 = 100$ km/s/Mpc.

\begin{figure*}
\centering
\includegraphics[width=17cm]{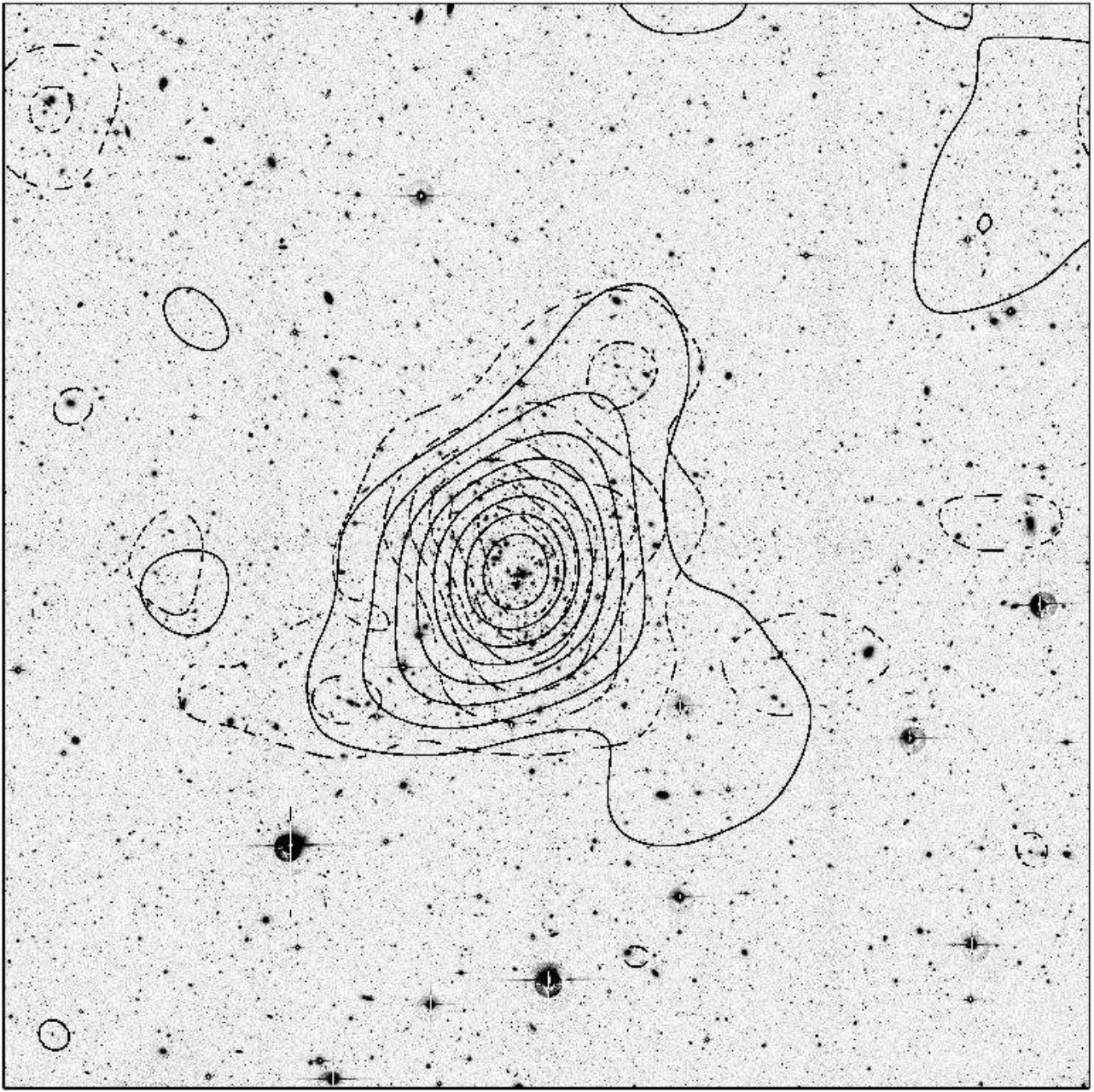}
\caption{Above is a $33\arcmin \times 33\arcmin$ R-band image of the cluster
\object{A1689} from the Wide Field Imager on the ESO/MPG 2.2m telescope plotted using
a $\sqrt{\log }$ stretch.  The two-dimensional mass reconstruction from
the weak lensing shear signal is drawn in solid-line contours.  The input 
shear field was smoothed using a $\sigma = 1\farcm 2$ Gaussian, which is 
roughly the smoothing level of the output mass reconstruction.  Each contour 
represents an increase in $\kappa $ of 0.015 above the mean $\kappa $ at the 
edge of the field.  The dashed-line contours show the flux-weighted
distribution of galaxies with $16.7 < R < 22$ also smoothed by a 
$\sigma = 1\farcm 2$ Gaussian.}
\label{fig1}
\end{figure*}

\section{Observations and Data Reduction}
Twelve 900 second exposures in $R$-band were obtained with the Wide Field
Imager (WFI) on the ESO/MPG 2.2m telescope on the night of May 29, 2000.
The images were taken with a dithering pattern between exposures which
filled in the gaps between the chips in the CCD mosaic.  The resulting image
covers a $34\farcm 7\times33\farcm 4$ area centered on the cluster with
$82.5\% $ of the area having the full exposure time and the remaining
$17.5\% $ of the area receiving less than the full exposure time due to the
area being out of the field of view or in the gap between the chips
during some of the exposures.  The final image has a FWHM on unsaturated
stars of $0\farcs 72$ and {a $1\sigma$ sky noise of 28.1 mag/sq arcsec on
the areas having the full exposure time.  Object counts at a $3\sigma $ 
detection limit in the regions containing the full exposure time are complete 
to $R=24.9$ for $2\arcsec$ radius aperture magnitudes, as measured by the 
point where the number counts depart from a power law. }

During the image reduction process, each chip in the CCD mosaic was treated
separately for the steps detailed below.  We first de-biased the images using
a master bias taken at the beginning of the night and corrected for bias
drift in each exposure using the overscan strip.  The images were then
flattened by a polynomial fit to a twilight flat taken during the evening
twilight.  The polynomial fitting was done using a 9th by 17th order
two-dimensional polynomial $f(x,y) = \sum_{l=0}^{9}\sum_{m=0}^{17} a_{lm}
x^l y^m$ with $a_{lm} = 0$ if $(l/9)^2 + (m/17)^2 > 1$ to accurately mimic
the large-scale variations in quantum efficiency while removing the small
scale variations in the twilight flat caused by fringing.  The ratio of powers
in the $x$ and $y$ directions were chosen so as to have the same density of
inflection points in the polynomial across the rectangular CCD.  A medianed
nightsky flat was then made from all of the long-exposure, twilight-flattened 
$R$-band images taken that night.  The nightsky flat showed a peak-to-peak 
fringing amplitude of $\sim 10\%$ ($[\mathrm{max}-\mathrm{min}]
/[0.5\times (\mathrm{max}+\mathrm{min})] \sim 0.1$ in adjacent minima and maxima).

We tried two different techniques to remove the fringing and flatfield the
images with the nightsky flat.  The first technique was to fit the nightsky
flat with a 9th by 17th order two-dimensional polynomial and flatten the images
with the polynomial.  The flattened images were then medianed to produce
another nightsky flat, which was assumed to contain only the fringing pattern.
This fringing pattern was then scaled to the sky level in each image and 
subtracted.  The second technique was to simply flatten the images with
the original nightsky flat, fringes included.  Neither technique is strictly
correct as in the former case we are subtracting small-scale changes in the
quantum efficiency (dust, chip defects, etc) instead of dividing by them,
while in the latter case we are treating the interference-based fringe
pattern as a variation in the quantum efficiency.  We used both sets of
flattened images to create two different final images and performed all
further analysis on both images.  In each case, the results for both images
were statistically identical (differences between the images were at least
one order of magnitude smaller than the $1\sigma $ error bars of the
measurement).  For the rest of the paper we discuss only the results from
the first technique image, but none of the conclusions change when using the
second technique image.

One problem with the WFI is that bright stars can have several reflection
rings, many of which are not centered on the star which causes them.
In this field there are 16 stars of sufficient brightness to cause noticeable
reflection rings on the final image.  Each star has two small rings,
outer radii $\approx 48\arcsec $, offset from each other by a few pixels.
These rings are typically a few arcseconds from the star radially
away from the center of the field, with the separation getting larger with
the star's distance from the field center.  The brightest five stars
also have a much fainter third ring, outer radius $\approx 92\arcsec$,
which is centered about twice as far from the star as the smaller rings
in the same radial direction.  These reflection rings have structure not
only in both radial and tangential directions from the center of the ring,
but also random structures similar to, and probably caused in part by the
removal of, the fringing in the sky.  As a result, we were unable to 
remove these rings while preserving the haloes of any objects inside them.
The regions containing the reflections rings were therefore removed from
all further analysis below.

The sky level of each image was determined by detecting minima in a smoothed
image, and fitting the minima with a 7th by 15th order two-dimensional
polynomial.  The resulting sky fit was subtracted from each image, which 
allowed the removal of extended haloes from saturated stars without affecting 
the profiles of individual galaxies.  

The next step in the image reduction was to move each image into a common
reference frame while simultaneously removing any distortion in the image
introduced by the telescope optics.  To do this we assumed that each CCD
of the mosaic could be mapped onto a common detector plane using only a
linear shift in the $x$ and $y$ directions and a rotation angle in the
$x$-$y$ plane, and that these mappings are the same for each image.  
In doing so we assume that the detectors are aligned well
vertically and thus no change in the platescale, distortion, etc.~occurs
across chip boundaries and that the chips do not move relative to each other
while the instrument is subjected to thermal variations and/or varying flexure 
from pointing to different parts of the sky.  We justify these assumption later
using analysis of the resulting PSF ellipticities.  We then used a bi-cubic
polynomial to map the detector plane for each exposure to a common reference
frame.  The parameters for the two sets of mapping were determined
simultaneously by minimizing the positional offsets of bright, but unsaturated,
stars among the various images and with the USNO star catalog.  A downhill
simplex minimization method was employed to minimize the 21 free parameters
in the individual CCDs to detector plane mapping, and at each step the
detector plane to common reference plane mapping parameters were determined 
for each image using LU decomposition of a $\chi ^2$ minimization matrix.
The resulting best fit mappings had a positional
rms difference of .06 pixels ($\sim 0\farcs 015$) among the images and
$0\farcs 55$ between the image positions and USNO star catalog positions.
No significant deviations from zero average positional differences were seen 
in any large area of any image.  We also attempted the mapping from detector
plane to common reference plane for each image using 5th and 7th order
two-dimensional polynomials, but the resulting rms dispersions in stellar
positions did not improve over the bi-cubic polynomial mapping.
A discussion of this technique in greater depth can be found elsewhere
\citep{CL01.1}.  

Each CCD was then mapped onto the common reference frame using a triangular
method with linear interpolation which preserves surface brightness and
has been shown to not induce systematic changes in the second brightness
moments of objects in case of a fractional pixel shift \citep{CL00.1}.
The images were then averaged using a sigma-clipping algorithm to remove
cosmic rays.  The final image is shown in Fig.~\ref{fig1}.

\section{Lensing Analysis}
The first step in performing weak lensing analysis is detecting and measuring
the second moments of the surface brightness of faint background galaxies.
This was done using the IMCAT software package written by Nick Kaiser 
({\tt http://www.ifa.hawaii.edu/$\sim$kaiser/imcat}).  In addition to
obtaining the centroid position and second moments of the surface brightness 
of all detected objects, the software package also measures a Gaussian
smoothing radius at which the object achieves maximum significance from the
sky background (hereafter $r_\mathrm{g}$ and $\nu $ respectively), the average level
and slope of the sky around each object, an aperture magnitude and 
flux, and the radius which encloses half of the aperture flux ($r_\mathrm{h}$).
The software also calculates the shear and smear polarizability tensors 
($P_\mathrm{sh}$ and $P_\mathrm{sm}$) which define how the object reacts to an applied shear
or convolution with a small anisotropic kernel respectively (\citealt{KA95.4},
hereafter KSB, corrections in \citealt{HO98.1}).

Unsaturated stars brighter than $R=24$ were selected by their half-light
radius and used to model the PSF.  The half-light radius,
ellipticity, and the shear and smear polarizabilities of the stars varied
systematically across the field.  The half-light radius and the trace of the 
shear and smear polarizabilities across the entire field were fit well with 
a two-dimensional fifth-order polynomial, while the two ellipticity components 
were fit best with two-dimensional seventh-order polynomials.

It is important to note that in both the final combined frame and each input
image neither the best fit polynomials for the stellar
ellipticities nor the residual ellipticities of the stars after subtracting
the fit show any indication of a sudden change in the ellipticity across CCD
boundaries.  Because each CCD has three degrees of freedom with respect to the
vertical level of the focal plane, some areas on each chip will be somewhat 
out-of-focus.  While
this will result in some smooth variation of stellar half-light radii and
ellipticities over a single chip, it can also result in sudden changes in
both values across a chip boundary \citep{KA98.1}.  That no such
distortions occur across chip boundaries in this data implies that the
chips in the WFI are sufficiently aligned vertically that they are sampling
the same depth in the focal plane.  This also means that there should not be
sudden change in plate-scale across the chip boundaries, which allows us to
use just a linear transformation between individual chip coordinates and the
common detector plane coordinates as discussed above.  Finally, because of the
$\approx 100$ pixel gap between chips in the mosaic, a similar sudden change
in the PSF can occur in the gap areas using dithered exposures which have
different PSFs.  All twelve of the individual exposures coadded in this
data set, however, have sufficiently similar stellar ellipticities and half-light
radii that no change in the final PSF can be detected in the regions missing
one or more of exposures being coadded.

The ellipticities of the galaxies were corrected using 
$\vec{e}_\mathrm{c} = \vec{e}_\mathrm{o} - (\mathrm{tr} \mathcal{P}_\mathrm{sm}^*)^{-1} \mathcal{P}_\mathrm{sm}
\vec{e}_\mathrm{f}^* $ (KSB) where 
$\vec{e}_\mathrm{f}$ is the fitted stellar ellipticity field evaluated at the position 
of the galaxy, $\vec{e}_\mathrm{o}$ is the original measured ellipticity of the galaxy,
and $\mathrm{tr} \mathcal{P}_\mathrm{sm}^*$ is the fitted trace of the stellar smear 
polarizability evaluated at the position of the galaxy.  The effects of 
circular smearing by the PSF can then be removed from the galaxies using 
$\vec{g} = (\mathcal{P}_{\gamma})^{-1} \vec{e}_\mathrm{c}$ \citep{LU97.1}
where $\mathcal{P}_{\gamma} = \mathcal{P}_\mathrm{sh} - 
\mathcal{P}_\mathrm{sm} \mathrm{tr} \mathcal{P}_\mathrm{sh}^* (\mathrm{tr} \mathcal{P}_\mathrm{sm}^*)^{-1}$,
for which the $\mathrm{tr} \mathcal{P}^*$ denote the fitted traces of the stellar
shear and smear polarizability evaluated at the position of each galaxy.
The resulting $\vec{g}$'s are then a direct estimate of the reduced
shear $\vec{g} = \vec{\gamma}/(1-\kappa)$, where both the shear $\gamma $
and convergence $\kappa $, the dimensionless mass density, are second 
derivatives of the gravitational potential.  Because the measured 
$\mathcal{P}_{\gamma}$ values are greatly affected by noise, we fit
$\mathcal{P}_{\gamma}$ as a function of $r_\mathrm{g}$, and 
$\vec{e}_\mathrm{c}$.  Because the PSF size varied slightly over the image,
we divided the background galaxies into 4 bins based on the stellar 
$r_\mathrm{h}$ in their vicinity, and did the $\mathcal{P}_{\gamma}$ fitting
separately for each bin.
Simulations have shown that this technique reproduces the level of the
observed shear to better than one percent accuracy \citep{VA00.1,ER01.1,BA01.1}.

We choose for our catalog of background galaxies those with magnitudes between
$R = 23$ and $R = 25.5$  which have a maximum signal-to-noise over the object,
as detected by the peak finding algorithm, greater than 9.  From this
sub-sample we removed objects with raw (uncorrected for PSF smearing)
ellipticities greater than 0.5, half-light radii similar to or smaller than
stellar, sky backgrounds greater than $3\sigma$ of the mean for objects in
the catalog, sky background slopes greater than $3\sigma $ of the mean, or
with a shear estimate greater than 2. We also manually removed from the 
catalog any sources which appeared to be a superposition of two or more 
objects.  {The final catalog had 27,288 objects with a density of
25.9 objects/sq arcmin.}

\begin{figure}
\resizebox{\hsize}{!}{\includegraphics{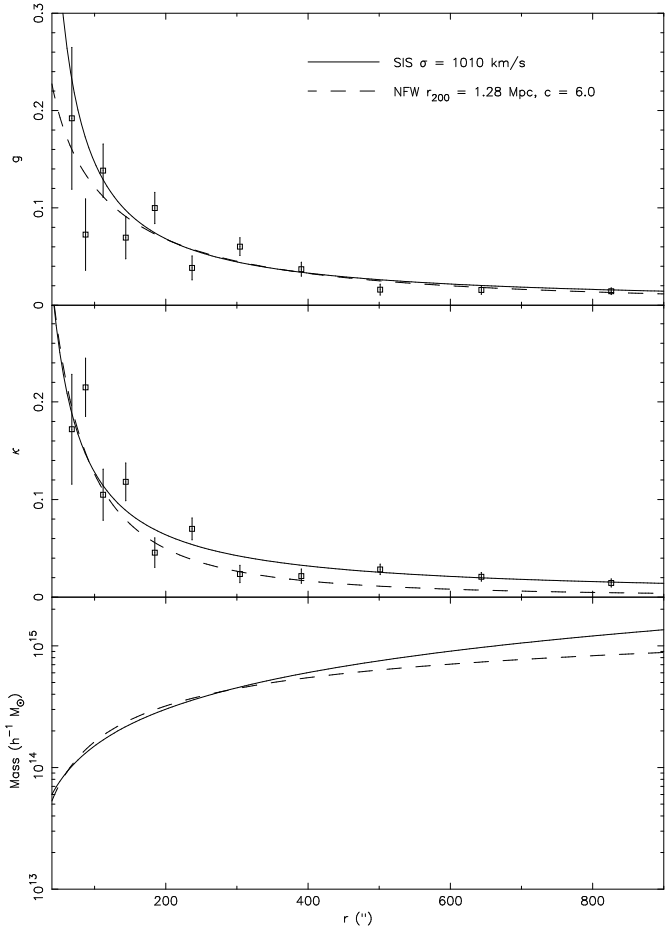}}
\caption{In the top panel above is plotted the reduced shear profile, 
radially averaged about the BCG, with $1\sigma$ error measurements for
each bin.  Also shown are the best fit SIS (solid line) and NFW (dashed line)
models.  In the middle panel are plotted the values for $\kappa $ derived
from the shear using Eq.~\ref{eq1} assuming the shear outside the $15\farcm 6$
outer radius follows an $r^{-1}$ profile.  Also shown are the $\kappa $
values for the SIS and NFW models given in the top panel.  The large radii
data points are reduced in $\kappa $ if a slope steeper than $r^{-1}$ is
assumed for the shear profile outside the outer radius, and agree well with
the NFW model for logarithmic slopes $\approx -1.7$.  The smaller radii data
points ($r < 400\arcsec$) are not greatly affected by changes in the slope.
The bottom panel shows the mass profiles of the SIS and NFW models given
in the top panel.}
\label{fig2}
\end{figure}

\begin{figure}
\resizebox{\hsize}{!}{\includegraphics{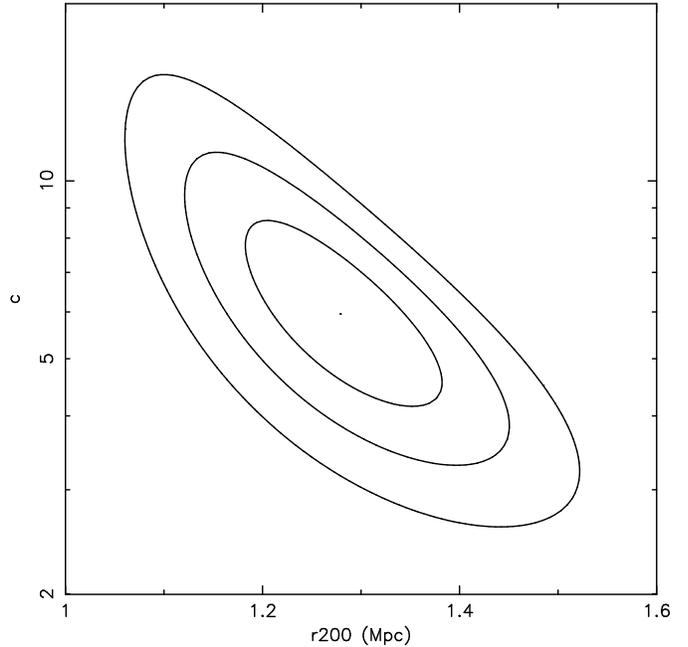}}
\caption{Shown above are the confidence contours for the NFW fit to the 
radial shear shown in Fig.~\ref{fig2}.  The plotted contours are for one,
two, and three $\sigma$ confidence levels (68.3\%, 95.4\%, and 99.73\% 
respectively) as measured by the change in $\chi ^2$ from the best fit 
model.}
\label{fig3}
\end{figure}

\begin{table}
\caption{\label{table1} Best fit model parameters for various background 
galaxy redshifts}
\begin{tabular}{cccc} \hline
 & SIS & \multicolumn{2}{c}{NFW} \\
$z_{\mathrm{bg}}$ & $\sigma $ (km/s) & $r_{200}$ ($h^{-1}$ Mpc) & $c$ \\ \hline
0.3 & $1475^{+40}_{-52}$ & 1.73 & 8.1 \\
0.5 & $1162^{+38}_{-42}$ & 1.42 & 6.7 \\
1.0 & $1028^{+34}_{-36}$ & 1.28 & 6.0 \\
3.0 & $962^{+33}_{-34}$ & 1.21 & 5.7 \\ \hline
\end{tabular}
\end{table}

The resulting background galaxy catalog was then used to measure the 
gravitational shear present in the field.  In Fig.~\ref{fig1}, overlayed in
solid contours on the R-band image, is the resulting massmap from a 
noise-filtering reconstruction \citep{SE96.3}.  As can be seen, there 
is a strong detection of the cluster
mass, the centroid of which is coincident with the position of the brightest
cluster galaxy.  {The flux-weighted distribution of galaxies with magnitudes
$16.7 < R < 22$ is also shown in Fig.~\ref{fig1} overlayed in dashed contours.
As can be seen, the distribution of bright galaxies has a shape very similar
to the mass distribution, with both having extended wings to the north-west
and south-east of the cluster.  The galaxy distribution does, however, have a
greater extension to the north-east in the cluster core than is seen in the
mass reconstruction.  Because we do not have color information on the galaxies,
we cannot distinguish which galaxies are cluster members and which could be
unrelated galaxies in projection.}

In Fig.~\ref{fig2} we plot the azimuthally-averaged reduced tangential shear 
using the position of the brightest cluster galaxy as the center of the 
cluster.  As can be seen in the figure, we have an observable shear signal
from $1\arcmin < r < 15\farcm 6$, and the shear in the outer bin, which
extends from $13\farcm 8 < r < 15\farcm 6$, is significant at greater than 
$4\sigma $.  Our shear profile agrees well with that of \citet{KA96.1} over
the $1\arcmin <r<5\arcmin $ aperture common to both data sets.  {Recent
work by \citet{KS01.1} has shown that cluster substructure, both in terms
of departures from sphericity of the cluster mass on the whole and putting
some of the mass in discrete galaxy haloes, has little effect on the
radial shear profiles and their best fit parameters.}

The radial shear profile is fit well by both a singular
isothermal sphere and a ``universal CDM profile'' 
\citep[][hereafter NFW]{NA97.6}, with both profile's best fit models having a 
$13.4\sigma $ significance from a zero mass model.  As can be seen in Table \ref{table1}, 
the mass of the best fit profiles are slightly dependent on the assumed mean redshift 
of the background galaxies used to measure the shear.  The models plotted in
Fig.~\ref{fig2} assume $z_{\mathrm{bg}} = 1$, but the significance of the
models do not change with the assume background galaxy redshift.
{Using the same magnitude cuts on a HDF-South photometric redshift catalog
\citep{FO99.1} as were used in the background galaxy catalog results in
a mean galaxy redshift of 1.15.  The average redshift, however, is measured
from only 48 galaxies, and thus is highly uncertain from both Poissonian noise
and cosmic variance. }  The parameters of
the best fit models were not significantly changed by either increasing
the inner or reducing the outer limiting radius of the fit.  {Using an 
F-test \citep{BE92.1}
to compare the 1 parameter SIS model and the 2 parameter NFW profile results
in the NFW providing a better fit with 95.5\% confidence.}  The errors on the
shear were calculated by measuring the rms dispersion of the ellipticity
component of the background galaxies $45\deg $ to the tangent with the cluster 
center (0.30 in this case) and dividing this by the square root of the 
number of galaxies in each bin.

Also in Fig.~\ref{fig2}, we plot $\kappa $ as a function of radius, as measured
using
\begin{equation}
1-\kappa (r) = {1\over 1-g(r)} \exp \left( -\int\limits^{r_0}_{r} {2 g(x)\over
x [1 - g(x)]} dx - {2 g(r_0)\over \alpha}\right)
\label{eq1}
\end{equation}
which assumes that at $r>r_0$ the reduced shear behaves as a power-law,
$g(r) \propto r^{-\alpha}$, and that $\lim_{r\to\infty} g = 
\lim_{r\to\infty} \kappa = 0$ \citep{SC95.1}.  For the values
plotted in Fig.~\ref{fig2}, we assume that the SIS profile continues at large radii,
and thus $\alpha = 1$.  If we assume instead that there is no mass outside
the limits in Fig.~\ref{fig2}, then $\alpha = 2$ and the resulting profile has
$\kappa$ at roughly one-quarter the value shown in Fig.~\ref{fig2} at large radii,
but nearly identical $\kappa$ at small radii.  This effect can also be seen
in the masses of the best fitting SIS and NFW models, in that while they
measure similar masses at small radii, at the edge of the field the NFW
model, which has a steeper slope in the reduced shear outside the image,
has a lower total mass than the SIS model.  

\section{Discussion}

We have shown in Section 3 that we have detected a weak lensing signal
at high significance centered on the BCG of \object{A1689}.  The best fit SIS mass
model, however, has a velocity dispersion of $\sim 1028^{+34}_{-36}$ km/s, which is
significantly below that measured using other techniques.  While one
could try to explain the higher mass estimates from X-ray emission and
cluster galaxy dynamics by invoking shock-heated gas and extended spatial
structure respectively, it is much harder to reconcile the weak lensing
shear measurement with the large Einstein radius obtained from strong lensing.
Because lensing measures the integrated mass along the line of sight, or more
precisely the integrated $\rho (z)/\Sigma_\mathrm{crit}(z)$, any model for the strong
lensing which involves multiple mass components at similar redshifts would
result in a weak lensing signal with a mass profile equal to the sum
of the individual model mass profiles.  The 1400 km/s and 700 km/s
dual isothermal sphere strong lensing model would be detected as a
1560 km/s isothermal sphere outside of a few arcminutes from the cluster
core.  

One possible method to have the large Einstein radius with the
lower mass weak lensing signal is to have two clusters causing the
strong lensing with the second at high redshift ($z>0.6$).  As the background
galaxies used in the weak lensing analysis are probably at a redshift
not much larger than the higher-redshift cluster, the second cluster would not 
contribute greatly to the weak lensing signal.  This model, however, has
two problems.  The first is that the strong lensing arcs would need to be
at fairly high redshift ($z>1.5$) for the high-redshift cluster to 
significantly
contribute to the strong lensing, and the relatively high surface brightness
of the arcs in the \object{A1689} system would suggest a lower redshift.  Second, the
high-redshift cluster would itself be lensed by the low-redshift cluster,
which, even if the clusters were arranged so that none of the high-redshift
cluster galaxies were strongly lensed by the low-redshift cluster, would
result in an overdensity of faint galaxies immediately outside the
Einstein radius of the low-redshift cluster.  We do not detect any such
overdensity of faint galaxies immediately outside the observed cluster core.

We know, however, that the mass profiles given in Section 3 are lower limits
to the true mass.  Because we have only a single passband for the field,
we were only able to select the background galaxy catalog on the basis of
magnitude, size, and significance.  As such, the catalog contains not only
the high-redshift galaxies we use to measure the weak lensing signal, but also
dwarf galaxies from both the cluster and foreground populations.  As a result,
one would need to correct the observed signal for the fraction of galaxies
which are not at high-redshift ($z\ga0.3$ in this case).  We consider
three different correction methods below.

For the first correction method, we assume that the fraction of non-background
galaxies is constant over the field, and so we multiply the reduced shear
estimate in each bin by a constant factor.  In order to obtain a best
fit SIS velocity dispersion of 1560 km/s (at a radius $> 3\arcmin$), 
assuming the background galaxies
have an average $z = 1$, the multiplication factor for the
shear estimate would be $\approx 2.5$.  This would mean that only $\approx 
40\%$ of the galaxies in the catalog are actually background galaxies.
Obtaining a best fit SIS profile which has the Einstein radius at $45\arcsec $
(1470 km/s using $z_\mathrm{bg} = 1$) would require that only $\approx 44\%$ 
of the galaxies are background galaxies.  
{If the strong lensing arcs are at higher redshift, then a slightly lower 
cluster mass is needed to make the $45\arcsec $ Einstein ring (1375 km/s
for $z_\mathrm{bg} = 3$) which means that $\approx 52\%$ of the galaxies in
the weak lensing catalog are background galaxies with $z\sim 1$.}
In order to have a $45\arcsec $ 
Einstein radius and not need negative mass just outside the Einstein radius 
to obtain the observed weak shear profile, one needs to have that at most 
$\approx 67\%$ of the observed galaxies are background galaxies.  

For the second correction method, we assume that the boost factor for each
bin is proportional to the mass in that bin.  This is equivalent to assuming
that the cluster dwarf galaxy population traces the mass at the outskirts
of the cluster, but falls off relative to the mass as one gets closer to
the core of the cluster.  To do this correction we first measure the reduced 
shear in each bin, and use Eq.~1 to determine $\kappa$ in the bin.  
Each bin's shear value was then multiplied by a constant times $\kappa$, where
the same constant was used for all bins, and a new $\kappa $ was calculated 
from the modified shear.
This was repeated until convergence.  In order to have a $45\arcsec $
Einstein radius without negative mass anywhere in the profile, the reduced
shears required a boost of $1.25 \times \kappa$, which resulted in the 
background galaxy/observed galaxy count ratio varying from $63\% $ at
$1\arcmin $ radius to $98\% $ at $14\arcmin $ radius.  In order to have 
a profile where $\kappa $ continues to increase between the $1\arcmin $
minimum weak lensing radius and the $45\arcsec $ Einstein radius, one
must use a minimum boost factor of $3.8 \times \kappa$, which results in a
background galaxy fraction which varies from $30\% $ at $1\arcmin $ radius
to $95\% $ at $14\arcmin $ radius.  It should be noted, however, that there
is not any boost factor using this technique which results in a SIS or NFW
profile which provides a good fit to both the resulting reduced shear
profile and the $45\arcsec $ Einstein radius.  We also tried assuming that
the dwarf galaxy population traces the mass everywhere in the cluster, but
this results in several bins consisting of only cluster galaxies.

\begin{figure}
\resizebox{\hsize}{!}{\includegraphics{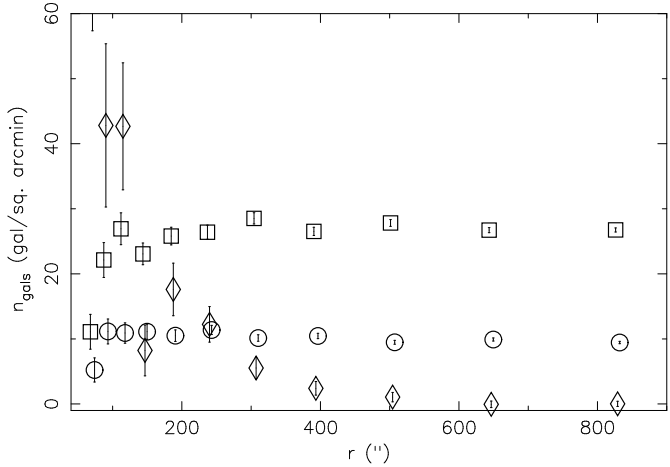}}
\caption{Plotted above are the number densities of galaxies in the background
galaxy catalog (open squares) and in the cluster galaxy catalog (diamonds) 
adjusted to the fainter magnitudes using the cluster galaxy number count 
slopes of \citet{TR98.1}.  The background galaxy catalog using the $R=24.2$
faint magnitude cut are plotted with open circles.
The galaxy densities have been corrected for loss
of area in each bin due to brighter objects.  Also plotted are $1\sigma $ 
error bars calculated from Poissonian noise of the detected number of 
galaxies in each bin.  As galaxies tend to be clustered on small scales,
however, these errorbars are smaller than the true statistical noise
\citep{TA98.1}.}
\label{fig4}
\end{figure}

For the final correction method, we selected objects from the master catalog
which had $17.4<R<19.4$ and half-light radii larger than stars, which are
presumably a field galaxy population and cluster galaxy population with
$0.5 L_{\star} \la L \la 3 L_{\star}$ \citep{KO98.1}.  By assuming the galaxy density
at the edge of the field is that of the field galaxies across the image,
we calculated the galaxy density of the cluster galaxies in the same
radial bins shown in Fig.~\ref{fig2}.  Using the cluster galaxy luminosity
function of \citet{TR98.1}, we scaled these galaxy counts to the magnitudes
of the background catalog, assuming the incompleteness of the faint end
would be the same for the cluster galaxies as background galaxies and
the cluster dwarf ellipticals have the same $B-V$ color as the $L_{\star}$ 
cluster ellipticals.  In Fig.~\ref{fig3} we plot the detected galaxy density
as a function of radius in the $23<R<25.5$ catalog and the cluster galaxy
counts scaled to this magnitude range.  As can be seen, this would result
in all of the detected galaxies within $\approx 2\farcm 5$ of the BCG
being cluster galaxies, and the fraction dropping off with radius.

The detected galaxy number density, however, actually decreases slightly 
as the radius decreases towards the BCG.  Thus, any increase in the density
of cluster dwarf galaxy density towards the cluster core would result in
an even more substantial decrease in background galaxy number density
with decreasing radius.  Studies of galaxy counts in random fields
give a typical slope of $\mathrm{d}\log N/\mathrm{d}m = 0.334$ for $22<R<27$
\citep{HO98.2}.  In a magnitude-limited sample, this slope would result in
the competing effects of magnification and displacement of background galaxies 
nearly canceling, and making only a very small decrease in galaxy counts with 
increasing lensing strength \citep{BR95.1}.  However, as our faint end cutoff
of the galaxy sample is a signal-to-noise cut rather than a magnitude cut,
the depletion of the background galaxies towards the core of the cluster
should be somewhat stronger than that of a magnitude limited sample.  This
is due to the lensing preserving the surface brightness of the background
galaxies but increasing the area.  Thus, while the total luminosity of the
lensed galaxy is increased by a factor $\mu$, the signal-to-noise is increased
by only $\sqrt{\mu}$.  Therefore, a galaxy which would have been magnified past
a magnitude cut might not be included in a signal-to-noise cut.  

Applying
a model to this to measure the mass of the cluster from the depletion signal,
however, would require a knowledge not just of $\mathrm{d}\log N/\mathrm{d}m$,
but of $\mathrm{d}^2\log N/\mathrm{d}m \mathrm{d}a$, where $a$ is a measure of
the intrinsic size of a galaxy.  When we instead use a galaxy catalog with
$23<R<24.2$, for which we are magnitude limited on the faint cut, we no longer
see evidence of a depletion towards the core and instead have a small
increase in number density towards the center of the cluster.  If this small
increase is taken as faint cluster galaxies, however, it would result in a
correction factor to the shear measurement in the inner part of the cluster
of less than $10\%$.  From this we suggest that the large mass measured from
the depletion by \citet{TA98.1} might be in part due to the loss of faint
magnified galaxies from the catalog due to a signal-to-noise cut in the
detection algorithm, unless their faint magnitude cut was sufficiently bright 
so that galaxies slightly larger than those detected would still be above the 
minimum signal-to-noise detection criteria.

One argument against the large field-wide correction factor needed to
increase the shear signal to the strong-lensing mass models is that in a 
photometric redshift survey based on the VLT deep
images of the HDF-S, \citet{FO99.1} find that of the 73 objects 
they detect with $23<R<25.5$, 60 are galaxies with $z>0.3$, one is a galaxy with 
$z\approx 0.05$, and the remaining twelve objects are faint stars.  As we
were unable to distinguish stars from faint galaxies based on their half-light
radii for $R>24$, eight of these stars would have been included our object
catalog, giving a background galaxy fraction of $87\pm5\%$.  \object{A1689} is at a
slightly higher galactic latitude than the HDF-S, so the faint stellar
fraction from the HDF-S should not be a severe underestimate of that of
the \object{A1689} field.  Correcting the reduced shear profile for an $87\%$
background galaxy fraction results in a best fitting SIS velocity
dispersion $\sigma = 1095^{+37}_{-38}$ km/s ($z_\mathrm{bg} = 1$ assumed).

Further, other weak lensing studies using both a similar magnitude range
for background galaxies and similar techniques for deriving the shear
field from the background galaxy ellipticities have measured masses for
clusters in agreement with both X-ray and dynamical mass measurements
\citep{SQ96.1,SQ96.2,SQ97.1}.  Weak lensing studies of high 
redshift clusters have found that either the majority of galaxies with
similar magnitudes to those used here have redshifts beyond 1, or 
\object{MS1054.4$-$0321}, \object{MS1137$+$6625}, and \object{RXJ1716$+$6708}, 
all at $z\approx 0.8$,
are the most massive clusters known \citep{LU97.1,CL00.1}.  
While color selection was used to select only
blue galaxies for the lensing analysis of the high redshift clusters, the
selection removed fewer than $20\%$ of the faint galaxies in the catalog.

{In summary, we have measured a shear signal around A1689 with the ESO/MPG WFI
between 1\arcmin and 15\arcmin from the BCG.  This shear profile is fit well
by both a 1030 km/s SIS model and a $r_{200} = 1.28 \:\mathrm{Mpc}, c = 6$ NFW model.
If we assume that 87\% of the faint objects in our catalogs are galaxies
with $z>0.3$ then the best fit mass model increases to a 1095 km/s velocity
dispersion.
Both models have masses well below the mass measured from the radius
of the strong lensing arcs (1375-1560 km/s depending on the model and
redshift of the arcs), but in agreement with the masses
measured from X-ray observations (1000-1400 km/s).  The velocity dispersion
of the cluster galaxies has been measured from 560 km/s to 2355 km/s depending
on how much of the redshift dispersion is caused by spatially extended
substructure along the line of sight.}

In order to resolve the discrepancy between the weak and strong lensing
masses, we suggest the following needs to be done:  First, spectroscopic
redshifts should be obtained for all of the strong lensing arcs visible
in the deep HST image (HST Proposal 6004) and a detailed strong lensing
model using the position, redshift, and shape information of the arcs
allowing for substructure both in the cluster and along the line of sight
needs to be made.  Second, wide field imaging of the cluster in multiple
passbands, both optical and infra-red, should be performed.  This will allow 
determination of photometric redshifts for all objects in a given 
magnitude range, and thus allow one to remove foreground stars and cluster
dwarfs from the lensing catalog.  Finally, multi-color deep imaging around
the cluster core should be done with HST to increase the number counts of
usable background galaxies immediately outside the strong lensing radius
and allow the measurement of the mass profile in the transition region between
strong and weak lensing with the lowest possible noise.

{Finally, we have run simulations to determine how many background
galaxies will be needed to distinguish NFW and SIS models with large
radii weak lensing shear profiles.  Using input assumptions of the galaxy
rms ellipticity of 0.31, number density of 25 galaxies/sq arcmin, cluster 
redshift of 0.186, background galaxy redshift of 1, and a profile radius
range of 1\arcmin to 15\arcmin, we
find that on average a cluster with a NFW density profile with $r_{200} = 
1.28 \:\mathrm{Mpc}, c = 6$ profile will be able to be distinguished from an SIS
model using an F-test with 94\% confidence (with the 95.5\% confidence we
measured within the dispersion of results for a single realization).  If three
such clusters were stacked, then the two models could be distinguished with
99.65\% confidence, and 99.9998\% confidence if ten such clusters were stacked.
If the outer radius of the profile is increased to 30\arcmin, then a single
cluster's models could be distinguished at 98.3\% confidence and three stacked
clusters best fit models at 99.98\% confidence.  If the inner 4\arcmin were
to be imaged to allow a number density of 100 galaxies/sq arcmin and the
profile minimum radius decreased to $0\farcm 66$, then with a single cluster
one could distinguish the NFW and SIS best fit models with 99.7\% confidence,
and 99.9999\% confidence with three stacked clusters.  If the input model is
an SIS, then the NFW and SIS best fit models are statistically
indistinguishable over these radii, although the best fit are generally not
in agreement with the typical profiles seen in N-body simulations 
\citep{NA97.6}. }

\begin{acknowledgements}
We wish to thank and acknowledge Lindsay King and Neil Trentham for help
and useful discussions.  We also wish to thank the referee, Genevieve Soucail,
for her comments which improved the quality of the paper.  
This work was supported by the TMR Network
``Gravitational Lensing: New Constraints on Cosmology and the
Distribution of Dark Matter'' of the EC under contract
No. ERBFMRX-CT97-0172.

\end{acknowledgements}

\bibliographystyle{apj}
\bibliography{H2933.bib}

\end{document}